\documentclass{ws-procs9x6}
\usepackage{overpic}
\usepackage{epsfig}

\newcommand{\CK}{\v Cerenkov}

\begin{document}
\vspace{-4.cm}
\title{AMS --- a magnetic spectrometer on the international space station}
   
\author{
\vspace{-0.5cm}
Lu\'isa Arruda, F. Bar\~ao, G. Barreira, J. Borges, F. Carmo,\\
  P. Gon\c{c}alves, Rui Pereira, M. Pimenta}
\address{LIP/IST ---
         Av. Elias Garcia, 14, 1$^o$ andar ---
         1000-149 Lisboa, Portugal \\
         e-mail: luisa@lip.pt, pereira@lip.pt}

\maketitle

\vspace{-0.5cm}

\abstracts{
The Alpha Magnetic Spectrometer (AMS) is a particle detector, designed to search
for cosmic antimatter and dark matter and to study the elemental and isotopic
composition of primary cosmic rays, that will be installed on the International
Space Station (ISS) in 2008 to operate for at least three years.
The detector will be equipped with a ring imaging \CK\
detector (RICH) enabling measurements of particle electric charge and
velocity with unprecedented accuracy. Physics prospects and test beam results
are shortly presented.} 
\vspace{-1.2cm}

\section{The AMS experiment}
The Alpha Magnetic Spectrometer (AMS)\cite{bib:ams} is a particle detector that
will be installed on the International Space Station (ISS) in 2008 and operate
for at least three years. A successful test of the concept was made with
an experimental version flight of AMS aboard the US
Space Shuttle Discovery for 10 days in June 1998.
AMS has a large geometrical acceptance ($\sim$0.5\,m$^2$.sr) and will be
equipped with a superconducting magnet to detect charged particles (up to
iron) in a large range of energy (from $MeV$ up to $TeV$) and to detect gamma
rays. The long exposure period of AMS in space will allow the accumulation of a large
statistics of events increasing in several orders of magnitude the sensitivity
of the proposed physical measurements.


\vspace{-0.3cm}

\section{Velocity and charge reconstruction in the RICH detector}
The inclusion of a ring imaging \CK\ detector (RICH) will provide AMS 02 
with additional and accurate measurements of particle velocity ($\beta\equiv v/c$) and
electric charge ($Z$).
The RICH is composed of a dual radiator (silica aerogel with $n=1.05$ and NaF),
a high reflectivity lateral conical mirror and a detection matrix with
photomultipliers coupled to light guides.
An electromagnetic cone of radiation with an aperture angle related to the
particle velocity ($\cos \theta_c=\frac{1}{n \beta}$) can be emitted
($\beta >\frac{1}{n}$) when the charged particle crosses the radiator material.
The particle direction ($\theta$, $\phi$) is reconstructed with high accuracy
from signals left on silicon planes. 
For the velocity reconstruction with the RICH, a maximum likelihood approach was
applied. The overall probability of the detected hits to belong to the expected
photon pattern 
is computed as $P(\theta_c)
= \prod_{i=1}^{n_{hits}} p_i^{n_{pe}} \{r_i(\varphi_i;\theta_c)\}$, where
$p_i$ is the hit probability evaluated from its distance to the pattern
($r_i$). The angle $\theta_c$ which maximizes the function $P(\theta_c)$
corresponds to the best estimation of the emission angle of electromagnetic radiation
(\CK\ angle).
%
%
%
%
%
Electric charge can be reconstructed from the number of radiated and detected
photons ($N_i$)
which is proportional to $Z^2$ and to the length ($L$) of radiator
crossed: $N_i \propto Z^2 \Delta L \left( 1 - \frac{1}{\beta^2 n^2}
\right)\epsilon_i$, where $\epsilon_i$ is essentially the ring acceptance.
\vspace{-0.3cm}
\section{Results with the RICH prototype}
\begin{figure}[tbb]
\vspace{-1.0cm}
\begin{center}
\begin{tabular}{ccccc}
\epsfig{file=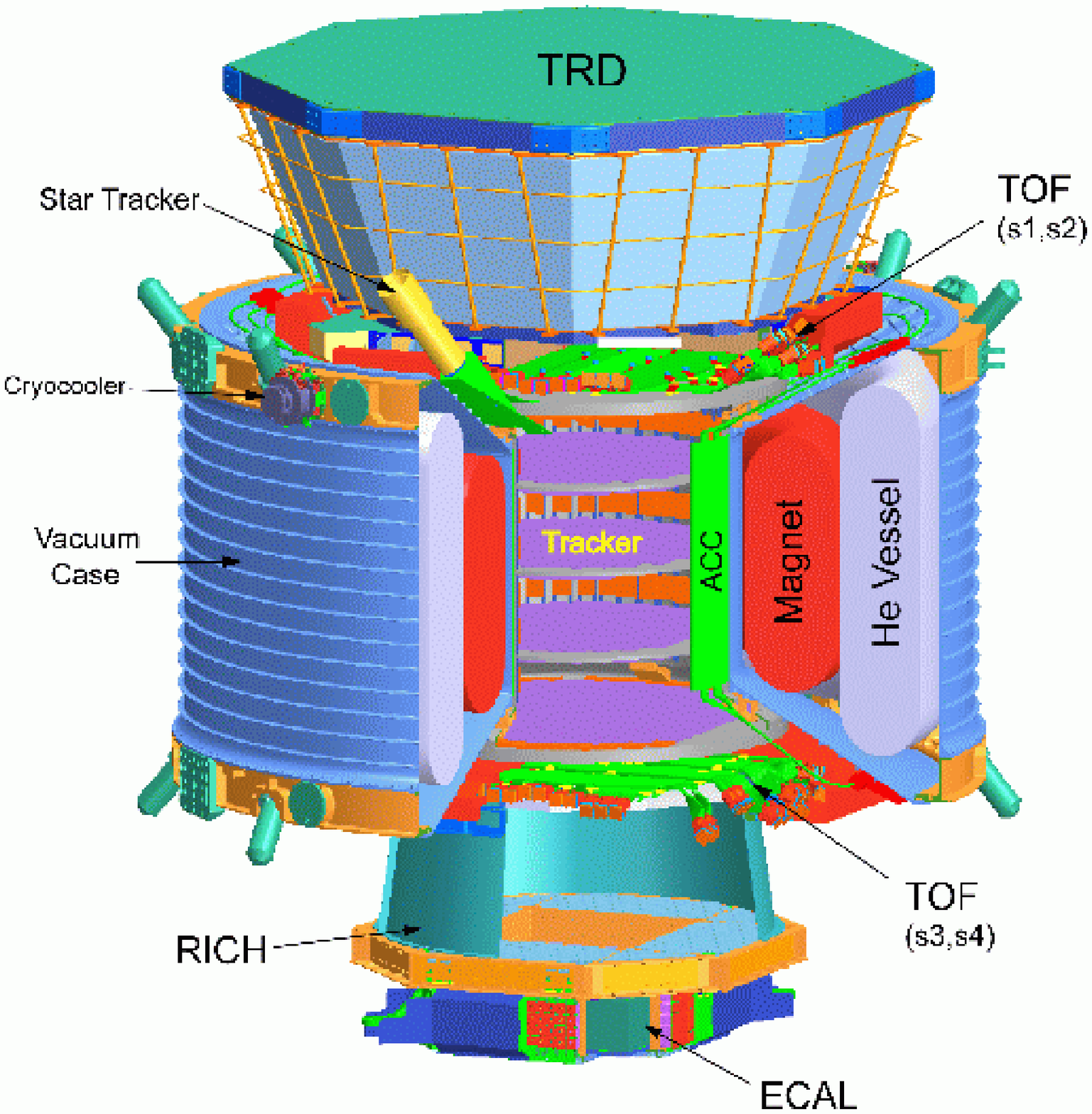,width=0.5\linewidth,clip=,bbllx=0,bblly=0,bburx=612,bbury=550} 
& & & &
\hspace{-1.3cm}
\scalebox{0.28}{%
\includegraphics[bb=-5 -5 485 507]{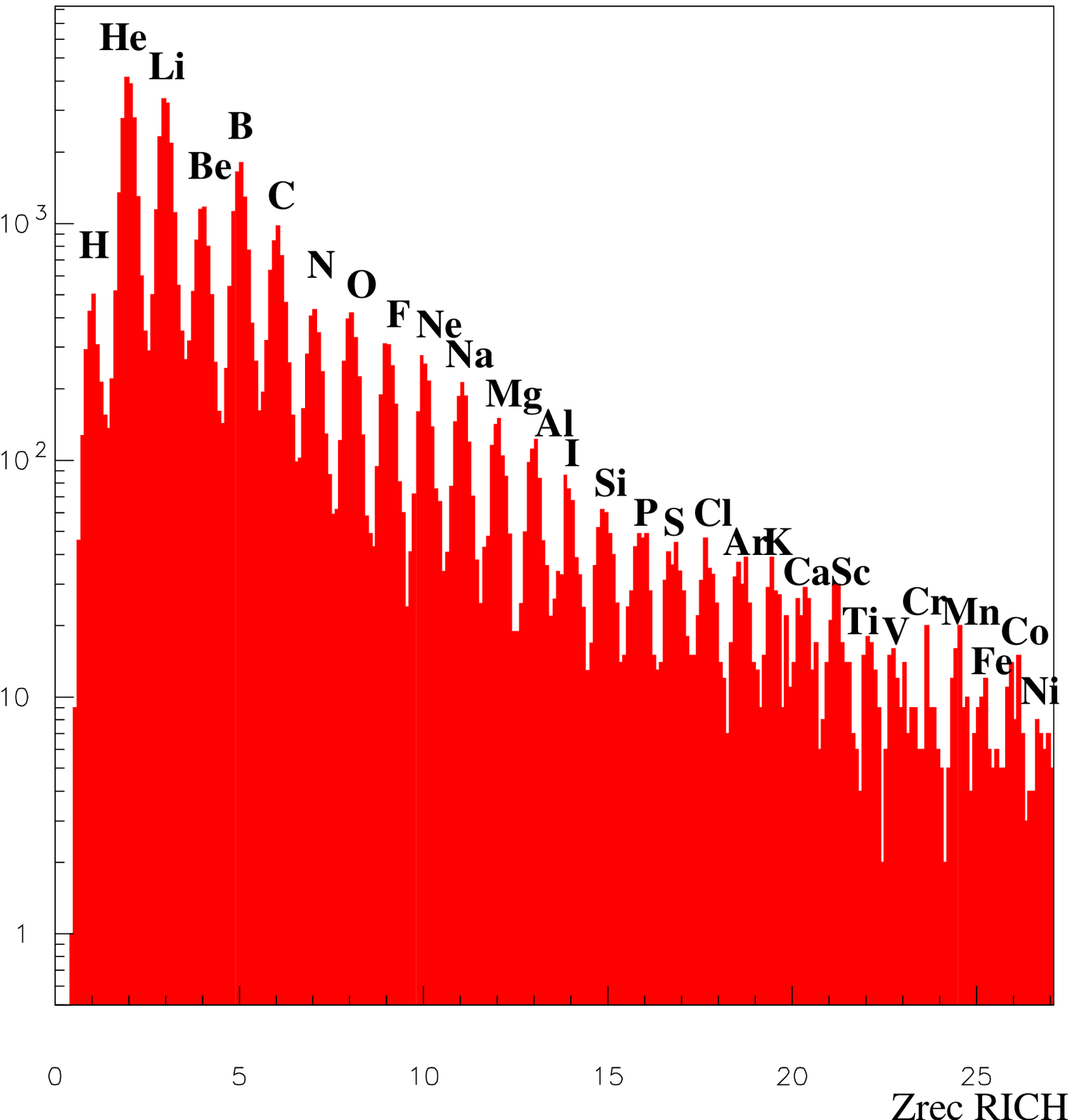} 
}                                                                          
\end{tabular} 
\vspace{-0.4cm}
\caption{At left a whole view of the AMS 02 spectrometer, and
  at right the reconstructed charge peaks using data from the RICH
  beam test at CERN in October 2003 with an indium beam of
  158\,GeV/nucleon.
\label{fig:thc}}
\end{center}
\vspace{-0.6cm}
\end{figure}                       

A RICH prototype made of a 96-photomultipliers was
tested with 158\,GeV/nucleon indium ion fragments at
CERN in 2003. Different types of radiators were tested as well as a reflector
segment. The collected data allowed to test the velocity and electric charge
reconstruction algorithms as well as the characterisation of the optical
properties of the radiators.

Figure \ref{fig:thc} shows the clear separation of nuclei up to iron element.
Velocity reconstruction with the same test beam data showed a velocity
resolution improving with the charge $Z$ as expected,
$\frac{\Delta\beta}{\beta}=(\frac{A}{Z})^2\oplus B^2$, with $A=7.8 \times
10^{-4}$ and $B=7.3 \times 10^{-5}$. The charged resolution obtained is
$\sim 0.2$ charge units with a systematic error of $\sim1\%$.


\vspace{-0.3cm}
\begin{thebibliography}{99}                                 
\bibitem{bib:ams}        S. P. Ahlen et al., Nucl. Instrum. Methods A {\bf 350}, 34 (1994).\\
                         V. M. Balebanov et al., {\em AMS proposal to DOE}, approved April 1995.
\bibitem{bib:NIM}{F. Barao et al., Nucl. Instrum. Methods A {\bf 502}, 310 (2003)}.
\end{thebibliography}
\end{document}